# Assessing the effectiveness of empirical calibration under different bias scenarios

Hon Hwang[1], Juan C Quiroz[1], Blanca Gallego[1*]

[1] Centre for Big Data Research in Health (CBDRH), University of New South Wales, Sydney, 2052, NSW, Australia

Corresponding author:

Blanca Gallego

Centre for Big Data Research in Health

Level 2, AGSM Building

G27, Botany St, Kensington NSW 2052

b.gallego@unsw.edu.au

**Key words:** Bias, confounding, empirical calibration, negative controls, positive controls, simulations

# Abstract

**Background:** Estimations of causal effects from observational data are subject to various sources of bias. One method of adjusting for the residual biases in the estimation of a treatment effect is through negative control outcomes, where the treatment does not affect the outcome. The empirical calibration procedure is a technique that uses negative controls to calibrate p-values. An extension of empirical calibration calibrates the coverage of the 95% confidence interval of a treatment effect estimate by using negative control outcomes as well as positive control outcomes (where treatment affects the outcome). Although empirical calibration has been used in several large observational studies, there is no systematic examination of its effect under different bias scenarios.

**Methods:** The effect of empirical calibration of confidence intervals was analyzed using simulated datasets with known treatment effects. The simulations consisted of binary treatment and binary outcome, with biases resulting from unmeasured confounder, model misspecification, measurement error, and lack of positivity. The performance of the empirical calibration was evaluated by determining the change in the coverage of the confidence interval and the bias in the treatment effect estimate.

**Results:** Empirical calibration increased coverage of the 95% confidence interval of the treatment effect estimate under most bias scenarios but was inconsistent in adjusting the bias in the treatment effect estimate. Empirical calibration of confidence intervals was most effective when adjusting for the unmeasured confounding bias. Suitable negative controls had a large impact on the adjustment made by empirical calibration, but small improvements in the coverage of the outcome of interest were also observable when using unsuitable negative controls.

**Conclusions:** This work adds evidence to the efficacy of empirical calibration of the confidence intervals in observational studies. Calibration of confidence intervals is most effective where there are biases due to unmeasured confounding. Further research is needed on the selection of suitable negative controls.

# 1 Introduction

Observational studies are often used when a randomised controlled trial design is unethical, costly, or time-consuming [1]. The trade-off is the loss of randomisation of treatment assignment, which is not guaranteed in observational studies. The lack of randomisation introduces confounding, where there are common causes for both the treatment and outcome [2, 3]. Confounding can lead to biases in the estimate of the treatment effect. Thus, treatment effect estimation in observational studies should include adjustments for confounders, i.e., using inverse probability score weighting.

Not all confounding can be accounted for. Residual confounding occurs when confounding variables are not measured, are measured incorrectly, or when the relationships between the confounders and the outcome are incorrectly modelled [2, 3]. One technique to account for residual confounding is through the use of *negative control* outcomes, which are treatment-outcome pairs where it is believed the treatment does not affect the outcome. Estimates of treatment effect from negative control outcomes can be used to adjust for the biases in the estimate of treatment effect on the outcome of interest, with the assumption that the negative controls and outcome of interest share the same casual structure [4]. Empirical calibration is a method that uses negative and, in an extension, positive controls, to adjust for the biases from residual confounding.

Initially, empirical calibration was proposed to calibrate the p-values of treatment effects through an empirical null distribution derived from negative controls [5]. Empirical calibration of p-values uses a Gaussian model of the negative controls to shift and scale the test statistics used to calculate p-values [6]. This idea was extended to empirical calibration of the confidence interval of an estimate of a treatment effect by using negative controls and *positive controls*—synthetically generated treatment-outcome pairs where the treatment has an effect on the outcome [7]. Empirical calibration of confidence intervals has been applied in several observational studies [8-12], where the calibration increased the coverage of the confidence intervals—to the "nominal" 95% coverage for 95% confidence intervals. While a prior study examined the limitations of empirical calibration of p-values [6], no study to date has assessed the effectiveness of empirical calibration of confidence intervals under different bias scenarios.

In this paper, we systematically examine the effect of the confidence interval by simulating different types of residual confounding. The simulations were carried out in the context of binary treatment and binary outcome with biases resulting from unmeasured confounder, model misspecification, measurement error, and lack of positivity. The simulations examined the effect of empirical calibration in terms of bias-variance trade-off for each type of bias. Our work has implications for observational studies estimating the comparative effectiveness of treatment strategies that plan to use empirical calibration to address residual confounding.

## 2 Methods

### 2.1 Empirical calibration procedure

Two types of treatment-outcome relationships are central to the empirical calibration procedure. In the calibration of p-values, an empirical null distribution is created using negative controls. To calibrate confidence intervals, both negative and positive controls are used to build an empirical *systematic error model*. Parameters from these empirical models are then incorporated into the calculation of p-values or confidence intervals.

#### 2.1.1 Negative Controls

A negative control outcome is an outcome not believed to be affected by the treatment of interest [4, 7, 13]. For example, Jackson et al. [14] used hospitalisation due to injury or trauma as a negative control outcome when examining the effect of influenza vaccination [14]. In this example, hospitalisation due to injury or trauma was not considered to be plausibly linked to an effect of influenza vaccination. Similarly, Schuemie et al. [7] used ingrown nail as a negative control outcome when comparing the adverse effect of using the drug dabigatran or warfarin on patients with atrial fibrillation; since dabigatran or warfarin were not considered to affect ingrown nail.

In empirical calibration of p-values, treatment estimates on negative controls are used to construct an empirical null distribution. Estimated parameters of the empirical null distribution are then incorporated into test statistics used for hypothesis testing. In practice, potential negative controls are identified from literature or existing negative and positive control reference sets [7].

To the extent possible, negative controls should have the same potential confounding mechanism as the outcome of interest (bias-comparable) [4, 6, 13]. In an ideal scenario, the common causes of the treatment and negative control outcome are identical to those of the treatment and outcome interest. In practice, this may not be possible because the treatment is likely to affect the various outcomes differently with varying magnitudes.

### 2.1.2 Positive Controls

Empirical calibration of confidence interval also relies on positive control outcomes, which are outcomes for which the treatment of interest has known effects. Unlike negative control outcomes, obtaining suitable positive controls is challenging. Even if there is a known positive control, an estimate of its effect size may be highly uncertain due to the study design [11]. If the effect size of a positive control is obtained from a randomised controlled trial, factors such as inclusion/exclusion criteria may not match that of the study with the outcome of interest. A secondary issue is that there may not be a wide range of positive effects available. This is because: (1) there tend to be insufficient target effect sizes for specific research contexts, (2) if target effect sizes are found, the magnitude of the effect size is unknown or it may be dependent on the population from which it was obtained [15].

To sidestep the challenges of obtaining a range of positive controls, synthetic positive control outcomes are generated from the negative controls [7]. Negative control outcomes are first modelled using penalised regression with incidence rate ratios as the unit for the treatment effect. From this model, subjects with the highest predicted probability of experiencing the outcome were then re-sampled and added to the treated group (in a binary treatment setup), resulting in a positive control with the desired treatment effect.

In this study, following previous simulation studies of bias adjustment [16, 17], the treatment effect was measured in terms of log odds ratio. We generated positive controls by reusing the estimated regression coefficients from negative controls and setting the treatment effect to an adjusted target log odds ratio. The adjustment took into account non-zero treatment effects from the negative controls indicative of potential biases [13]. For simplicity, we assumed a linear relationship between confounders and the logit of the control outcomes.

To calibrate confidence intervals, the empirical calibration procedure constructs a *systematic error model* using both types of controls. Similar to the calibration of p-values, parameters from the systematic error model are incorporated into the calculation of the confidence interval [7, 15].

## 2.2 Simulations

We conducted a set of simulation experiments to determine the effect of empirical calibration when bias is present due to four common sources [2]: unmeasured confounding, model misspecification (due to missing quadratic or interaction term), lack of positivity, and measurement error. For each bias type, the performance of empirical calibration was assessed under three different data generation conditions: (1) the limit ("ideal") case in which the negative controls share identical potential confounder effects as the outcome of interest; (2) confounders affect all outcomes (negative controls and outcome of interest) via beta parameters selected at random; (3) the other limit ("worse") case scenario in which the negative controls do not share the same potential causal pathway with the outcome of interest. Conditions (1) and (2) emulate cases where suitable negative controls are selected, i.e. they are bias-comparable to the outcome of interest. Case (3) emulates cases where unsuitable negative controls are selected, i.e. the negative controls are not bias-comparable to the outcome of interest.

The simulation process consisted of three steps: (1) generating a set of potential confounders X; (2) generating the binary treatment values $Z$; and (3) generating the binary outcomes consisting of the outcome of interest $Y^\star$ and $S$ negative controls $Y^-$. We assumed that the outcome of interest and the negative control outcomes are all measured as part of the same large observational dataset. For simplicity, and following previous literature [16, 18], we used logistic linear regression as our base treatment and outcome models.

### 2.2.1 Generating confounders

We included ten measured confounders ($m = 10$) sampled from a normalised Gaussian distribution $X_m \sim \mathcal{N}(\mu = 0, \sigma^2 = 1)$. Across the simulations, the number of confounders and the sample size were kept constant as the focus of the study was to assess the effect of empirical calibration under different bias inducing scenarios. Supplementary material 1 shows the directed acyclic graph of this simulation setup with one negative control outcome and two measured confounders.

### 2.2.2    Generating the treatment variable

The treatment assignment probability was modelled as a logistic function that depended on the linear combinations of the confounders parametrised by $\alpha = \{\alpha_0 + \alpha_1 + ... + \alpha_m\}$. Treatment values were then sampled from a Bernoulli distribution with probabilities:

$$\Pr(Z = 1 \mid X) = \text{logistic}(\alpha_0 + \alpha_1 X_1 + \alpha_2 X_2 + \cdots + \alpha_m X_m),$$

$$= \frac{1}{1 + \exp[-(\alpha_0 + \alpha_1 X_1 + \alpha_2 X_2 + \cdots + \alpha_m X_m)]} \ ,$$

$$Z \sim \text{Bern}[\Pr(Z = 1 \mid X)],$$

where $X = \{X_1, X_2, ..., X_m\}$.

### 2.2.3    Generating the outcome of interest and the negative controls

To generate the outcome of interest and negative controls, the probability of outcome $Y = 1$ was modelled as a logistic function that depends on confounders and treatment variable. The combination of confounders that the logistic function depends on varied depending on the types of bias we wanted to introduce. The reference logistic function is a linear combination of independent variables

$$\Pr(Y^\star = 1 \mid X) = \text{logistic}(\beta_0^\star + \beta_z^\star Z + \beta_1^\star X_1 + \beta_2^\star X_2 + \cdots + \beta_m^\star X_m) \ (6)$$

$$Y^\star \sim \text{Bern}[\Pr(Y^\star = 1 \mid X, Z)],$$

with $\{\beta_0^\star, \beta_1^\star, ..., \beta_m^\star\}$ as the regression parameters. Different bias scenarios were simulated by modifying this reference outcome model to include non-linear or interactive terms in equation 6.

The negative control outcome probabilities were also generated from a logistic model, but without the dependency on the treatment:

$$\Pr(Y_s^- = 1 \mid X) = \text{logistic}(\beta_{s0}^- + \beta_{s1}^- X_1 + \beta_{s2}^- X_2 + \cdots + \beta_{sm}^- X_m) \ (7)$$

$$Y_s^- \sim \text{Bern}[\Pr(Y_s^- = 1 \mid X)],$$

where $\{\beta_{s0}^-, \beta_{s1}^-, \beta_{s2}^-, \ldots, \beta_{sm}^-\}$ are the regression parameters for $s$-th negative control outcome. For brevity, $f_s^\star$ denotes the data generating function for the outcome of interest and $f_s^-$ denotes the data generating function (and outcome model) for the $s$-th negative control.

### 2.2.4 Simulating bias

The bias scenarios simulated in this study are summarised in Table 1: unmeasured confounding, model misspecification, lack of positivity, and measurement error. Each bias scenario consisted of 500 simulation iterations with $N = 50,000$ subjects, resulting in 500 comparisons between the calibrated and uncalibrated estimates of the treatment effect. The relatively large sample size of 50,000 was chosen in the reference scenario to avoid biases associated with sample size effects. Within each simulation iteration, the treatment, outcome, and control model parameters were randomly sampled from a uniform distribution extending from $\log(0.5) = -0.693$ to $\log(2) = 0.6931$.

*Unmeasured confounding*: Confounding not measured or not controlled for in the analysis [3]. This was introduced in our experiments by adding an extra confounder $U$ to the simulated outcome models $f^\star$ and $f_s-$, without making the confounder available at the time of treatment effect estimation. In one experiment (2.1) the negative controls shared the same unmeasured confounder as the outcome of interest. In the first instance, we simulated an ideal case in which the parameter associated with the unmeasured confounding in the outcome of interest ($f^\star$) is replicated in all negative control models $f_s-$. We then relaxed this constraint and allowed all the confounder parameters to be independently generated. In a second experiment (2.2), the extra confounder $U$ was added to the outcome of interest ($f^\star$), but not to the negative controls ($f_s-$).

*Model misspecification*: Biases associated with various forms of misspecification in the outcome models. In our experiments, we introduced a quadratic term $X_1^2$ (experiment group three) and an interaction term between confounders $X_1 X_2$ (experiment group four). In the first group of experiments (3.1 and 4.1), the negative controls shared the same misspecification as the outcome of interest, first with identical parameters and then with parameters chosen at random. In the second group of experiments (3.2 and 4.2), model misspecification was only present in the outcome of interest.

*Lack of positivity*: Lack of overlap between the treatment groups (in terms of propensity score), also known as structural violation of the positivity assumption. This was introduced in experiments 5.1 and 5.2 by modifying regions of the propensity score distribution as described in [19].

*Measurement error*: Confounders contain a systematic error or are subject to noise. This was introduced in experiment 6 by adding an error term $E \sim \mathcal{N}(\mu_\epsilon, \sigma_\epsilon^2)$ to the confounder with the largest effect size. The mean (greater than zero) and standard deviations were chosen at random for each simulation iteration.

### 2.2.5 Applying the empirical calibration procedure to the simulated data

The *systematic error model* for empirical calibration can be derived using regression estimates from (1) all the negative and positive controls or (2) only the negative controls (referred to as the null error model). The null error model assumes the systematic error is the same for all true effect sizes, whereas using all negative and positive controls models the systematic error as a function of the true effect size. Treatment effect estimates for the negative and positive controls are calculated using a standard inverse propensity score weighted logistic regression (referred to as estimation function). The systematic error model is then constructed using these estimates. The estimation function is also applied to the outcome of interest to estimate the treatment effect and its model-robust 'sandwich' standard error. This estimate is then calibrated using the systematic error model resulting in calibrated estimates and their corresponding standard errors. See Supplementary material 2 for additional details on modelling the systematic error in empirical calibration.

For brevity, we refer to empirical calibration with the systematic error model as "default empirical calibration." Sensitivity analysis included comparing default empirical calibration with empirical calibration with the null model. Five negative controls were used for empirical calibration, but sensitivity analysis included calibration with 30 negative controls to examine the effect of increasing the number of negative controls.

We used version R v4.0.3 [20] on the x86_64 architecture. Stabilised weights in the estimation function were obtained using version 0.10.0 of the WeightIt package [21], and their sandwich standard errors were obtained using the v4.0 survey package [22]. Calibration of the estimates was performed using version v2.1.0 of the EmpiricalCalibration package [7]. Our simulation software is licensed under GPLv3 and is available at https://github.com/clinical-ai/assess-empcalib/.

Funnel plots were used to represent the biases in the estimates of the treatment effect on the outcome of interest with and without calibration across each confounding scenario (**Table _1_**). For each experiment, the funnel plots were generated for default empirical calibration with five negative controls. The funnel plots also include the results from sensitivity analysis: (1) empirical calibration with the null model, and (2) empirical calibration with 30 negative controls.

# 3    Results

## 3.1   Calibration of treatment effect on the outcome of interest

Figure 1 to 5 show funnel plots representing the biases in the estimates of the treatment effect on the outcome of interest with and without calibration across the bias scenarios outlined in Table 1 (experiments 2 to 6). The corresponding differences in coverage by the 95% confidence interval and bias of the outcome of interest after applying empirical calibration are presented in Table 2. The average width of the confidence intervals after calibration are shown in **Table _4_**. Supplementary material 3 contains coverage plots of the calibrated controls in this study (presented as a diagnostic of empirical calibration performance in [7]). Running an experiment without any introduced confounding (experiment 1) resulted in zero bias (see Supplementary material).

Empirical calibration increased the coverage of the confidence intervals of the outcome of interest in 8/10 cases when negative controls were suitable (bias-comparable to the outcome of interest), with coverage increasing in the range of +1% to +60%. Empirical calibration increased coverage when bias was due to unmeasured confounding, model misspecification (quadratic term), non-positivity, and measurement error, though the increase for non-positivity (+2%) and measurement error (+1%) was small. When bad negative controls (not bias-comparable) were used for empirical calibration, coverage increased in 4/4 cases, though the increase in coverage was modest by comparison (+1% - 8%). The small increase in coverage for non-positivity and measurement error is due to uncalibrated confidence intervals having coverage of the outcome of interest close to 95% (Table 3). In bias-comparable scenarios where empirical calibration increased the coverage, the average width of the confidence intervals also increased (**Table *4***).

Empirical calibration decreased bias in the outcome of interest in 8/10 cases when negative controls were suitable, in the scenarios of unmeasured confounder, model misspecification (interaction term), and measurement error. For non-positivity and model misspecification (quadratic term), the effect in bias was inconsistent. When unsuitable negative controls were used, calibration decreased bias in 3/4 cases.

## 3.2   *Empirical Calibration with the Null Model*

Empirical calibration with the null model decreased coverage in 9/10 cases when compared with default empirical calibration when suitable negative controls were used (coverage decrease ranging from -1% to -52%). Calibration decreased coverage for model misspecification (quadratic and interaction term), non-positivity, and measurement error. For unmeasured confounding, the change in coverage was inconsistent (-6% and +9%). With unsuitable negative controls, coverage decreased in 4/4 cases when compared with default empirical calibration (-3% to -8%).

With suitable negative controls, empirical calibration with the null model increased bias in the outcome of interest in 8/10 cases when compared with default empirical calibration. Bias increased for model misspecification (quadratic and interaction term) and measurement error. For unmeasured confounder and non-positivity, the change in bias was inconsistent. With unsuitable negative controls, bias increased in 2/4 cases.

## 3.3   *Empirical Calibration with a Higher Number of Negative Controls*

Increasing the number of negative controls from five to 30 increased coverage in 7/10 cases when suitable negative controls were used, the increase ranging from +1% to +6%. Coverage increased for unmeasured confounder, model misspecification (interaction term), and non-positivity. When using unsuitable negative controls, coverage increased in 2/4 cases. Increasing the number of negative controls had an inconsistent effect on the bias, decreasing bias in 5/10 cases when suitable negative controls were used and decreasing bias in 2/4 cases when unsuitable negative controls were used.

# 4    Discussion

## *4.1  Main Findings*

This paper examined the impact of empirical calibration across different types of biases introduced in simulation scenarios. In the majority of the simulations, empirical calibration increased coverage of the 95% confidence interval and decreased the bias of the outcome of interest. Across the bias scenarios, empirical calibration performed best in bias due to unmeasured confounders. Both suitable and unsuitable negative controls resulted in increased coverage and decreased bias, though the increase in coverage was lower when unsuitable negative controls were used for calibration. This suggests that when suitable negative controls are chosen for calibration, it benefits both coverage and bias of the outcome of interest, but calibration with unsuitable negative controls provides small benefits.

While prior work discussed the assumption that the negative control outcome should share the same potential causal mechanism as the outcome of interest to be effective [4, 6, 13], our work demonstrates small gains even when less than ideal negative controls were used for empirical calibration. While the performance of empirical calibration was mostly positive, its performance was at times inconsistent in our bias scenarios in how it affected coverage and bias, with cases where coverage was increased at the expense of increased bias. This suggests further examination of empirical calibration in controlled simulated datasets and in real-world datasets, particularly in selecting appropriate controls.

Our results showed minimal gains from running empirical calibration with 30 negative controls instead of five negative controls. A previous study [11] used 76 negative controls, with each one obtained by examining literature. Depending on the domain, this may not be feasible, and it is time consuming. Our results confirm findings from prior work, showing gains when the number of negative controls is increased [7], but if a few negative controls are chosen with proper evidence, they can be sufficient for robust performance of empirical calibration.

Empirical calibration with the null model (using only negative controls) yielded lower coverage gains than default empirical calibration (using negative and positive controls). That is, the assumptions made by null model calibration decreased the effectiveness of empirical calibration. However, the small gains in coverage of the 95% confidence interval should encourage practitioners to apply empirical calibration with the null model in cases where it is unfeasible or too complex to derive positive controls from the negative controls.

A major issue with applying empirical calibration of the confidence interval is the suitability of negative controls. A higher number of negative controls increases the chance of having at least one negative control that shares the same causal structure as the outcome of interest. A null (zero) treatment effect estimate is necessary but insufficient for the negative control to be suitable, as capturing the bias from the same causal structure as the outcome of interest is what enables effective empirical calibration. Our results emphasize the importance of having suitable negative controls, as results show that empirical calibration of confidence interval works most effectively when using negative controls that share a similar causal structure as the outcome of interest. A prior study also identified this issue when analysing empirical calibration of p-values [6]. Given the importance of negative controls to the empirical calibration procedure, an important area of future work is investigating ways of selecting suitable negative controls which share the same causal structure as the outcome of interest.

Once negative controls are identified, empirical calibration software [7] can synthetically generate positive controls and perform the calibration of confidence intervals. Calibrated confidence intervals can be presented with noncalibrated intervals to enable readers to assess the uncertainty attributable to systematic error [7]. One method of assessing the impact of the empirical calibration in practice (where true treatment effects are not available) is to apply the calibration procedure to the negative and positive controls, comparing the changes to the null (zero) and synthetic positive treatment effects before and after calibration. This method is detailed in prior work [11].

There are alternative methods to adjust biased treatment effect estimates. For example, instrumental variables (IV) can be used as part of a sensitivity analysis to address potential unmeasured confounding [23]. Simulation-extrapolation (SIMEX) is a technique for biases introduced due to measurement error in covariates [24]. These techniques can be used in place of or in combination with empirical calibration. However, a comparison of these methods with empirical calibration is left for future work.

## 4.2 Limitations

Our simulations explored five common bias scenarios, but real-world data is likely to result in combinations of different types of biases. The data generation and estimation functions were linear models, with non-linear and non-parametric functions to be explored in future work. The confounders in our simulations were independent, and the number of measured confounders was fixed, which limited the data simulations explored. The treatment effect in our analysis was the log of an odds ratio (regression coefficient of logistic regression). Future work needs to analyse how other treatment effects, such as relative risk and the corresponding generation of positive controls, may affect the performance of empirical calibration.

## 5   Conclusions

Our work adds evidence to prior studies that empirical calibration can increase coverage of the 95% confidence interval and decrease bias of the outcome of interest. This is relevant for observational studies estimating the comparative effectiveness of treatments. Empirical calibration performed best when adjusting bias due to unmeasured confounder. Caution must be taken to select suitable negative controls, as unsuitable negative controls lessen the efficacy of empirical calibration.

# Declarations

**Ethics approval and consent to participate**

Not applicable.

**Consent for publication**

Not applicable.

**Availability of data and materials**

The datasets generated and analysed during the current study are available in the assessing empirical calibration repository, https://github.com/clinical-ai/assess-empcalib/.

**Competing interests**

The authors declare that they have no competing interests.

**Funding**

This work was supported by National Health and Medical Research Council, project Grant No. 1125414.

**Authors' contributions**

HH and BG designed the data simulations and experiments. HH coded the data simulations and empirical calibration procedure. HH, BG, and JQ analysed and interpreted the results. HH drafted the initial manuscript. All authors contributed to critical revisions of the manuscript. All authors approved the final draft.

**Acknowledgements**

The authors thank Luca Maestrini for helpful comments. This work includes results produced on the computational cluster Katana at UNSW Sydney.

**Table 1.** Bias scenarios with covariate dependencies in the data generation and estimation functions. For each bias scenario, two experiment groups explore the effect of suitable negative controls (having the same potential confounding mechanism as the outcome of interest) and unsuitable negative controls. The third column shows the covariate dependencies of the data generating functions: $f^{\star}$ for the outcome of interest and $f^{-}$ for the negative controls. $X$ is the set of measured confounders.

| Bias Scenario | Experiment Group | Outcome generation function |
|---|---|---|
| Reference case | 1 | $f^{\star}$ and $f^{-}$ depend on $X$ |
| Unmeasured confounder $(U)$ | 2.1 | $f^{\star}$ and $f^{-}$ depend on $X$ and $U$ |
| | 2.2 | Only $f^{\star}$ depends on $X$ and $U$ |
| Model misspecification: Quadratic term $(X_1^2)$ | 3.1 | $f^{\star}$ and $f^{-}$ depend on $X$ and $X_1^2$ |
| | 3.2 | Only $f^{\star}$ depends on $X$ and $X_1^2$ |
| Model misspecification: Interaction between two confounders $(X_1 X_2)$ | 4.1 | $f^{\star}$ and $f^{-}$ depend on $X$ and $X_1 X_2$ |
| | 4.2 | Only $f^{\star}$ depends on $X$ and $X_1 X_2$ |
| Lack of positivity | 5.1 | $f^{\star}$ and $f^{-}$ depend on $X$. |
| | 5.2 | Only $f^{\star}$ depends on $X$. |
| Lack of overlap between treatment groups in terms of propensity score. | | |
| Measurement error in confounder | 6 | $f^{\star}$ and $f^{-}$ depend on $X$. <br><br> One of the confounders with large effect have random measurement error. |

**Table 2.** Differences in coverage of the outcome of interest by the 95% confidence interval and difference in mean of standardised absolute biases of the outcome of interest. Default empirical calibration was performed with five negative controls and using the all-controls systematic error model.

| Difference Coverage | | | | Difference in Mean of Standardised Absolute Bias | | |
|---|---|---|---|---|---|---|
| | Ideal—Suitable negative controls | Random—Suitable negative controls | Random—Unsuitable negative controls | Ideal—Suitable negative controls | Random—Suitable negative controls | Random—Unsuitable negative controls |
| **No calibration vs default empirical calibration** | | | | | | |
| Unmeasured confounder | 0.48 | 0.60 | 0.05 | -0.46 | -0.23 | 0.00 |
| Model misspecification Quadratic term | 0.24 | 0.27 | 0.08 | -0.05 | 0.03 | -0.01 |
| Model misspecification Interaction term | -0.09 | 0.04 | 0.02 | -0.19 | -0.09 | -0.02 |
| Non-positivity | 0.02 | 0.00 | 0.01 | -0.02 | 0.07 | -0.01 |
| Measurement error | 0.01 | 0.01 | NA | -0.02 | -0.02 | NA |
| **Default empirical calibration vs empirical calibration with the NULL systematic error model** | | | | | | |
| Unmeasured confounder | 0.09 | -0.06 | -0.04 | 0.55 | -1.56 | 0.00 |
| Model misspecification Quadratic term | -0.09 | -0.13 | -0.08 | 0.20 | 0.35 | 0.02 |
| Model misspecification Interaction term | -0.52 | -0.06 | -0.03 | 0.29 | 0.00 | 0.01 |
| Non-positivity | -0.02 | -0.01 | -0.03 | -0.01 | 0.09 | -0.01 |

| Measurement error | -0.04 | -0.02 | NA | 0.02 | 0.01 | NA |
|---|---|---|---|---|---|---|
| **Default empirical calibration (5 negative controls) vs empirical calibration with 30 negative controls** | | | | | | |
| Unmeasured confounder | 0.05 | 0.06 | 0.01 | 0.85 | -8.31 | -0.36 |
| Model misspecification Quadratic term | 0.00 | 0.00 | -0.05 | 0.04 | -0.11 | 0.08 |
| Model misspecification Interaction term | 0.02 | 0.02 | 0.02 | 0.06 | 0.06 | -0.60 |
| Non-positivity | 0.02 | 0.03 | 0.00 | -0.02 | 0.43 | 0.12 |
| Measurement error | 0.01 | -0.01 | NA | -0.11 | -0.06 | NA |

**Table 3**. Coverage of the outcome of interest by the 95% confidence interval without calibration and with empirical calibration.

| | Ideal—Suitable negative controls | | Random—Suitable negative controls | | Random—Unsuitable negative controls | |
|---|---|---|---|---|---|---|
| | Calibrated | Uncalibrated | Calibrated | Uncalibrated | Calibrated | Uncalibrated |
| Unmeasured Confounder | 0.79 | 0.31 | 0.91 | 0.31 | 0.36 | 0.31 |
| Model misspecification Quadratic term | 0.99 | 0.75 | 0.99 | 0.72 | 0.82 | 0.74 |
| Model misspecification Interaction term | 0.87 | 0.96 | 0.98 | 0.94 | 0.96 | 0.94 |
| Non-positivity | 0.96 | 0.94 | 0.94 | 0.94 | 0.96 | 0.95 |
| Measurement error | 0.96 | 0.95 | 0.96 | 0.95 | NA | NA |

**Table 4** Average width of 95% confidence intervals with and without calibration for reference simulation (five negative controls, random Gaussian noise for measurement error).

| | Ideal—Suitable negative controls | | Random—Suitable negative controls | | Random—Unsuitable negative controls | |
|---|---|---|---|---|---|---|
| | Calibrated | Uncalibrated | Calibrated | Uncalibrated | Calibrated | Uncalibrated |
| Unmeasured Confounder | 0.44 | 0.10 | 0.50 | 0.10 | 0.13 | 0.10 |
| Model misspecification Quadratic term | 0.61 | 0.11 | 0.62 | 0.11 | 0.13 | 0.11 |
| Model misspecification Interaction term | 0.39 | 0.11 | 0.44 | 0.11 | 0.13 | 0.11 |
| Non-positivity | 0.22 | 0.18 | 0.21 | 0.18 | 0.22 | 0.18 |
| Measurement error | 0.18 | 0.15 | 0.17 | 0.15 | NA | NA |

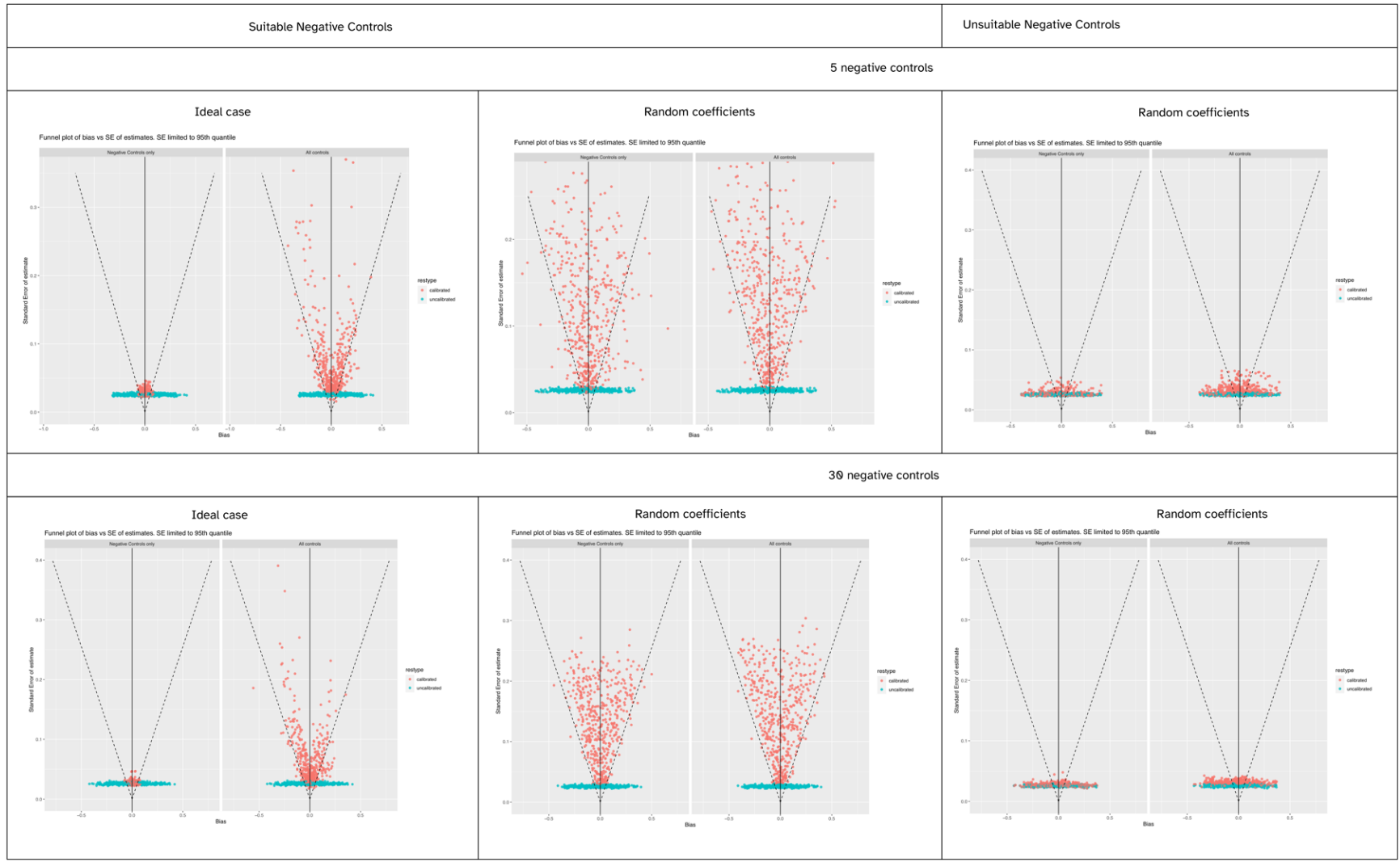

**Figure 1.** Bias of treatment estimates and the coverage of their standard errors by the 95% confidence interval in the unmeasured confounding scenario. In the "ideal" case the effect of confounders to the outcomes are the same, and in the "random coefficients" case these effects are randomised. Within each cell, the left funnel plot shows the estimates calibrated with negative controls only and the right funnel plot shows estimates calibrated with negative and positive controls.

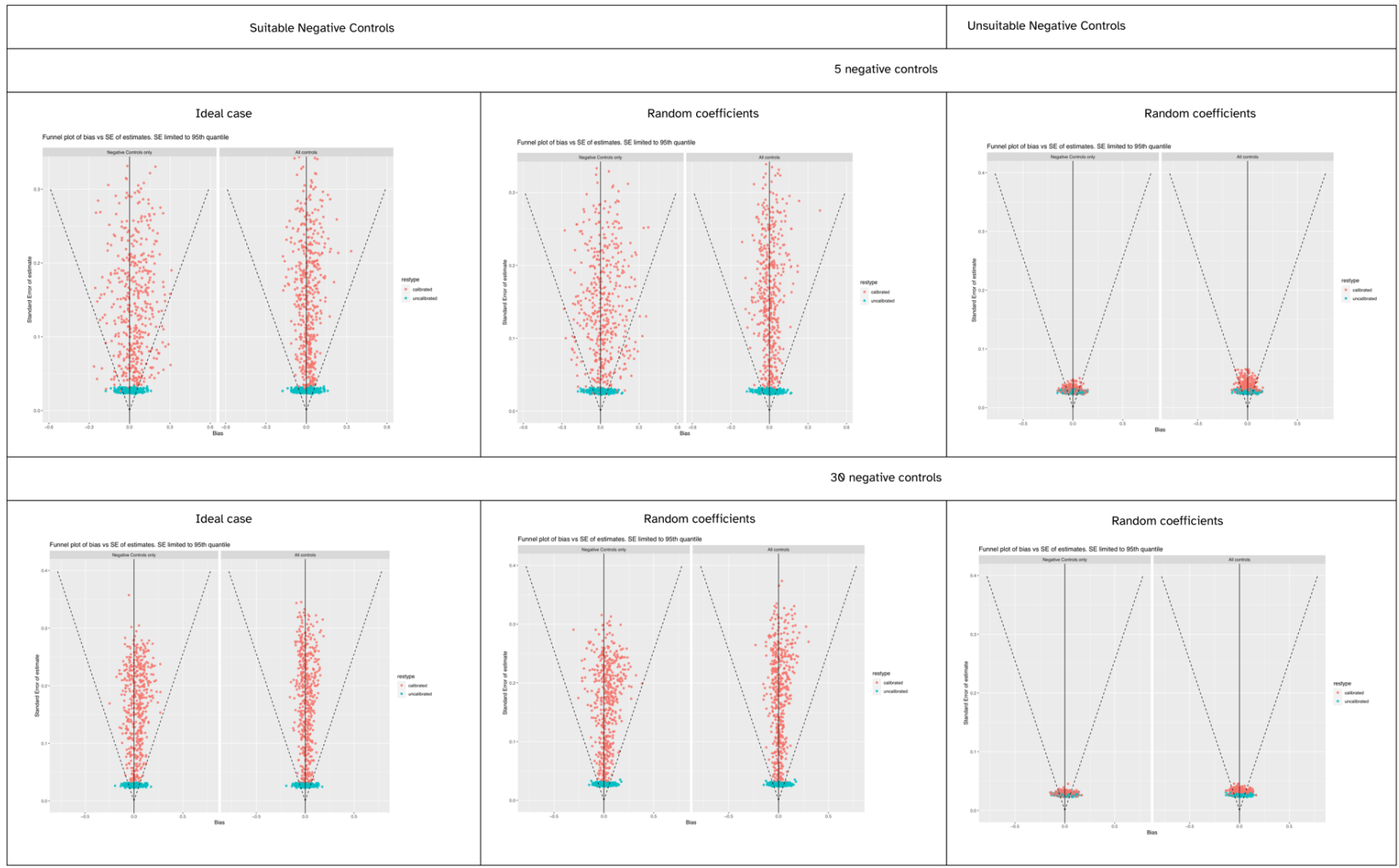

**Figure 2.** Bias of treatment estimates and the coverage of their standard errors by the 95% confidence interval in the model misspecification – missing quadratic term scenario. In the "ideal" case the effect of confounders to the outcomes are the same, and in the "random coefficients" case these effects are randomised. Within each cell, the left funnel plot shows the estimates calibrated with negative controls only and the right funnel plot shows estimates calibrated with negative and positive controls.

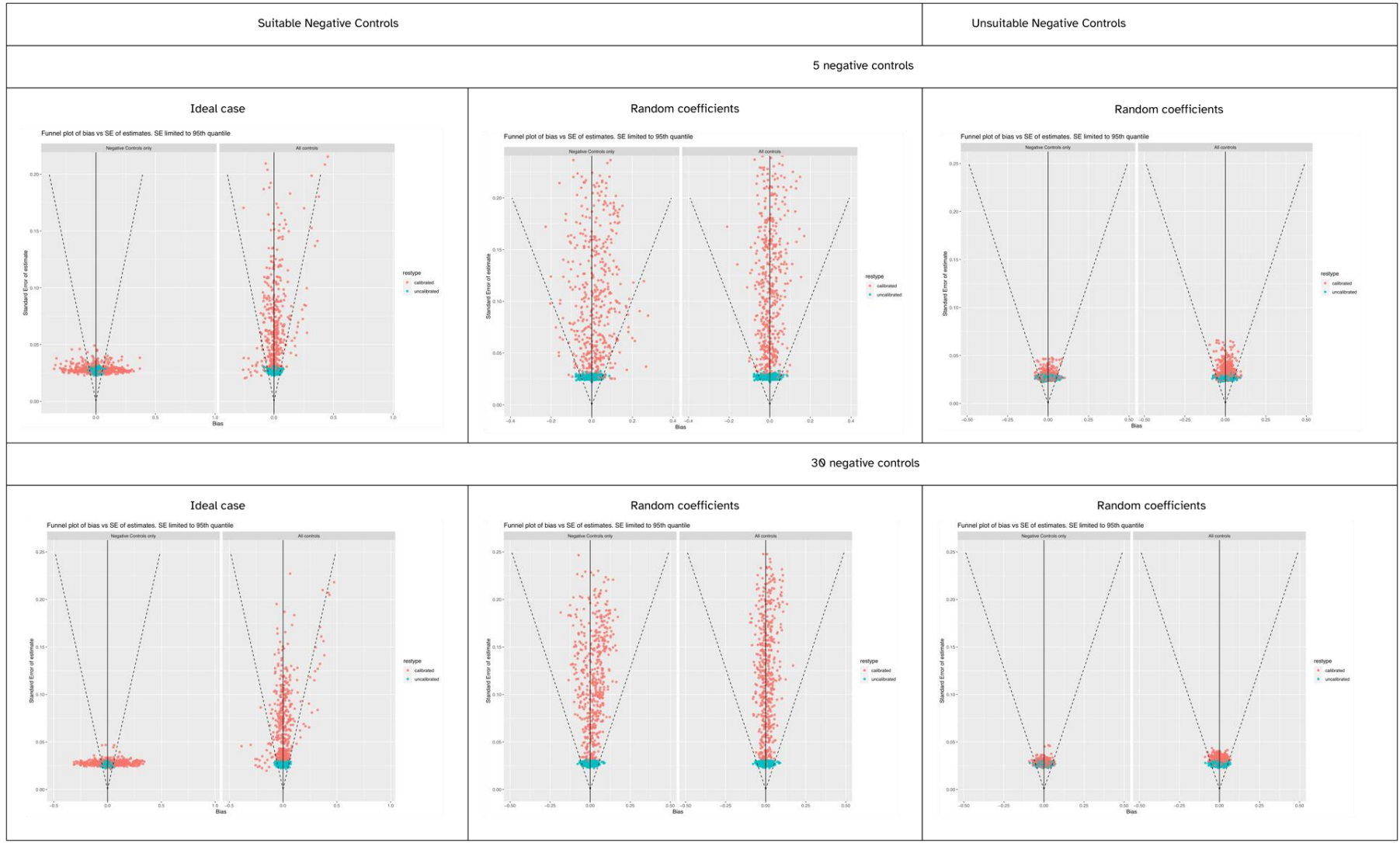

**Figure 3.** Bias of treatment estimates and the coverage of their standard errors by the 95% confidence interval in the model misspecification – missing

interaction term scenario. In the "ideal" case the effect of confounders to the outcomes are the same, and in the "random coefficients" case these effects are randomised. Within each cell, the left funnel plot shows the estimates calibrated with negative controls only and the right funnel plot shows estimates calibrated with negative and positive controls.

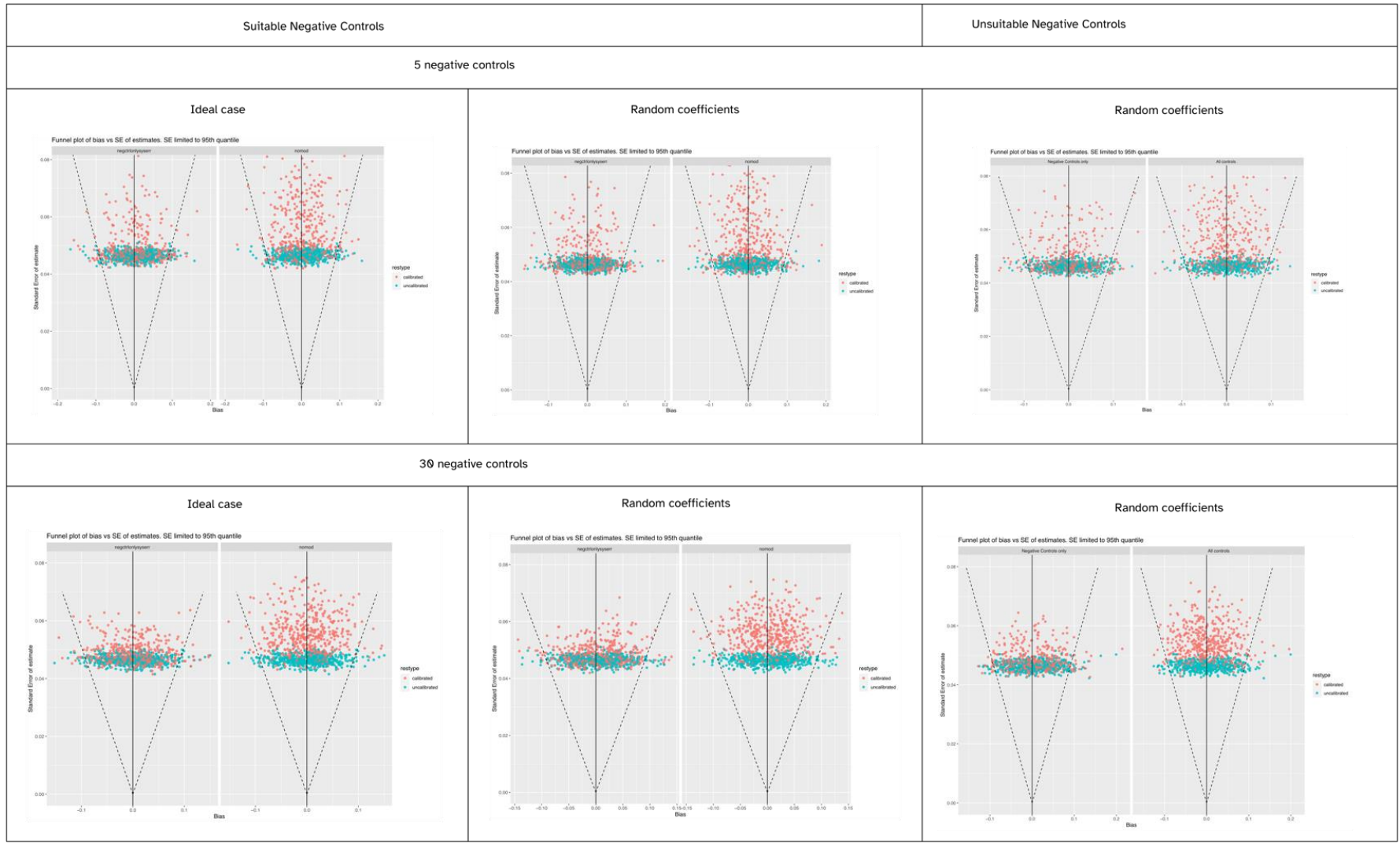

**Figure 4.** Bias of treatment estimates and the coverage of their standard errors by the 95% confidence interval in the non-positivity scenario. In the "ideal" case the effect of confounders to the outcomes are the same, and in the "random coefficients" case these effects are randomised. Within each cell, the left funnel plot shows the estimates calibrated with negative controls only and the right funnel plot shows estimates calibrated with negative and positive controls.

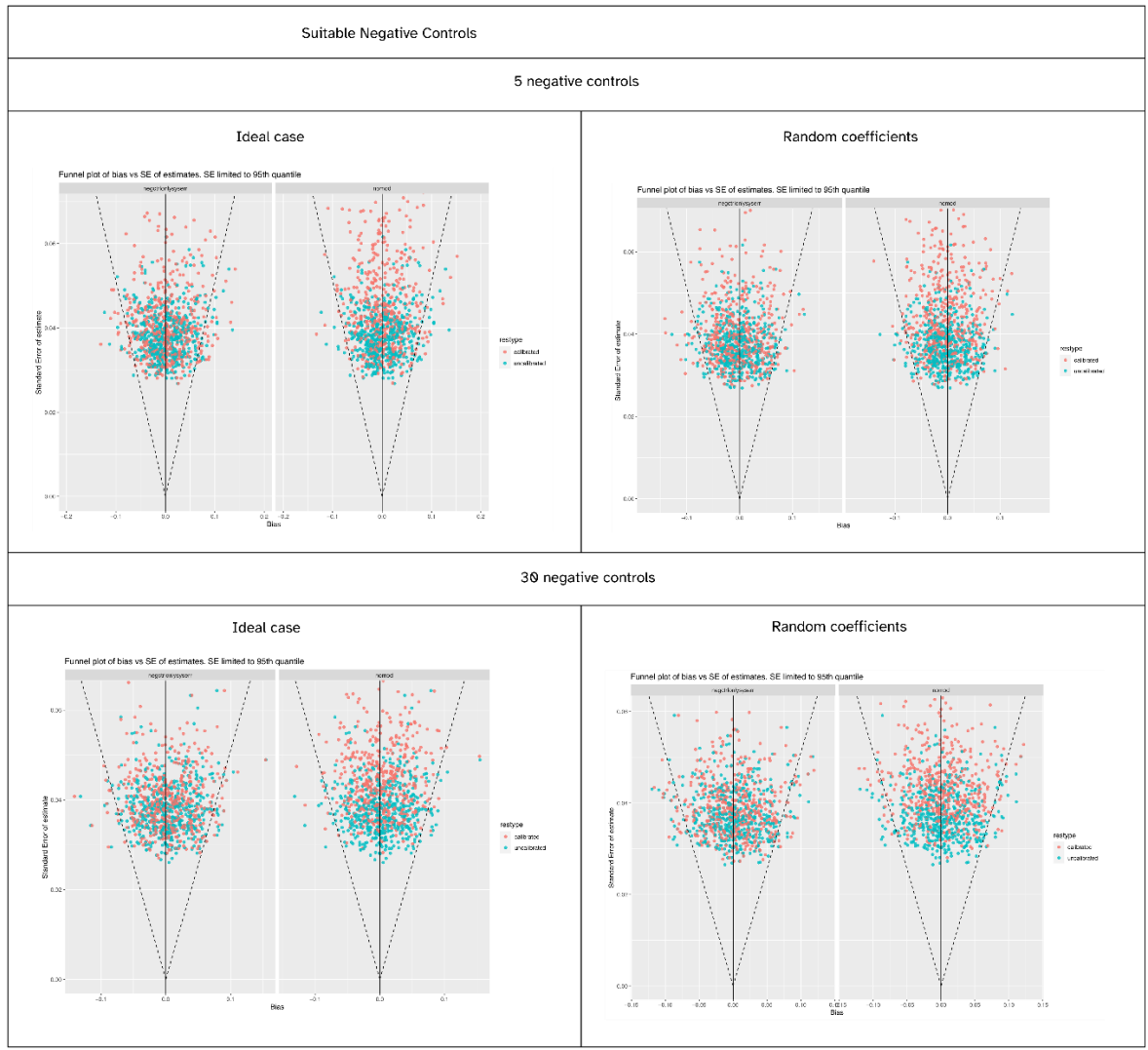

**Figure 5.** Bias of treatment estimates and the coverage of their standard errors by the 95% confidence interval in the measurement error scenario. In the "ideal" case the effect of confounders to the outcomes are the same, and in the "random coefficients" case these effects are randomised. Within each cell, the left funnel plot shows the estimates calibrated with negative controls only and the right funnel plot shows estimates calibrated with negative and positive controls.